\documentclass[12pt]{article}
\usepackage{amsmath,amsfonts}

\makeatletter\@addtoreset{equation}{section}\makeatother

\def\bH {\mathbb{H}}

\def\bZ {\mathbb{Z}}

\def\Z{\bZ}

\def\bC {\mathbb{C}}

\def\be{\begin{equation}}
\def\ee{\end{equation}}
\def\bea{\begin{eqnarray}}
\def\eea{\end{eqnarray}}

\newcommand{\m}{\mu}
\newcommand{\n}{\nu}
\newcommand{\p}{\nabla}

\newcommand{\cM}{{\mathcal M}}

\newcommand{\Tr}{{\rm Tr\,}}

\renewcommand{\title}[1]{\vbox{\center\LARGE{#1}}\vspace{5mm}}
\renewcommand{\author}[1]{\vbox{\center#1}\vspace{5mm}}
\newcommand{\address}[1]{\vbox{\center\em#1}}
\newcommand{\email}[1]{\vbox{\center\tt#1}\vspace{5mm}}

\begin{document}
\begin{titlepage}
\begin{center}
\hfill \\
\hfill \\
\vskip 1cm

\title{One-loop Partition Functions of 3D Gravity}

\author{Simone Giombi$^{1,a}$,
Alexander Maloney$^{2,b}$,
Xi Yin$^{1,c}$}

\address{$^1$Jefferson Physical Laboratory, Harvard University,\\
Cambridge, MA 02138 USA\\
\medskip
$^2$McGill Physics Department, 3600 rue University,\\
Montr{\'e}al, QC H3A 2T8, Canada}

\email{$^a$giombi@physics.harvard.edu,
$^b$maloney@physics.mcgill.ca,
$^c$xiyin@fas.harvard.edu}

\end{center}

\abstract{
We consider the one-loop partition function of free quantum field theory in locally Anti-de Sitter space-times.  In three dimensions, the one loop determinants for scalar, gauge and graviton excitations are computed explicitly using heat kernel techniques.  We obtain precisely the result anticipated by Brown and Henneaux: the partition function includes a sum over ``boundary excitations'' of AdS$_{3}$, which are the Virasoro descendants of empty Anti-de Sitter space.  This result also allows us to compute the one-loop corrections to the Euclidean action of the BTZ black hole as well its higher genus generalizations.
}

\vfill

\end{titlepage}

\eject \tableofcontents

\section{Introduction}

In this paper we present several new results on the one-loop
structure of quantum field theories in Anti-de Sitter space.
Our goal is to compute the Euclidean partition function
\be
Z = \int D\phi e^{-g^{-2}S(\phi)}
\label{Zpart}
\ee
of a free quantum field $\phi$ propagating in a fixed background $\cM$ which is locally Anti-de Sitter (AdS).  We have included an explicit factor of $g^{-2}$ (proportional to $1/\hbar$) in front of the action.
In this paper we will focus on the three dimensional case. As we are computing a Euclidean partition function, the metric of $\cM$ will be locally Euclidean AdS$_{3}$, which is 3-dimensional hyperbolic space $\bH_{3}$.  Any locally $\bH_{3}$ space $\cM$ is either $\bH_{3}$ itself, or a quotient of $\bH_{3}$ by some discrete group $\Gamma$.  For example, if we consider Eulcidean AdS$_{3}$ with the identification $t_{Euclidean}\sim t_{Euclidean}+\beta$, then (\ref{Zpart})
is the partition function of Thermal AdS.  Other examples include the Euclidean BTZ black hole and its higher genus generalizations.

We will consider the case where the field $\phi$ is either a scalar,
a $U(1)$ gauge field or a linearized metric perturbation.
The path integral (\ref{Zpart})
may be expanded around a classical solution $\phi_{0}$ to the equations of motion as
$$
\log Z = -g^{-2}S^{(0)}+S^{(1)}+g^{2}S^{(2)}+\dots
$$
Here $S^{(0)}=S(\phi_{0})$ is the action of the classical solution, and $S^{(i)}$ denotes the correction to this saddle point action at $i^{th}$ order in perturbation theory.  The goal of this paper is to compute the one loop action $S^{(1)}$ expanded around the classical vacuum solution $\phi=\phi_{0}$ for any locally hyperbolic space.
For gauge fields and metric perturbations this calculation is quite technical, although the end result -- described in sections 3, 4 and 5 below -- is relatively simple.

Perhaps the most interesting application of the results described above is the problem of three dimensional quantum gravity with a negative cosmological constant.  The Euclidean action of the theory is
\be
S = -{1\over 16 \pi G} \int  d^{3}x\sqrt{g} \left( R + {2\over \ell^{2}}\right)
\ee
where the length $\ell$ is related to the cosmological constant $\Lambda = -2/\ell^{2}$.  Solutions to the equations of motion are metrics of constant negative curvature $R=-6/\ell^{2}$.  The theory has a single dimensionless coupling constant, 
$k=\ell/16 G$. We will use units where $\ell=1$.

Our goal is to compute the partition function of quantum gravity with asymptotically AdS boundary conditions at a given temperature $\beta^{-1}$ and angular potential $\theta$. The canonical ensemble partition function at finite $\beta$ and $\theta$ can be thought of as the Euclidean functional integral
\be
Z(\tau) = \int_{\partial \cM=T^{2}} Dg e^{-k S(g)}
\label{Ztemp}
\ee
where we integrate over metrics whose conformal boundary a torus $T^{2}$ with modular parameter $\tau=\frac{1}{2\pi}(\theta+i\beta)$.  In writing (\ref{Ztemp}) we have pulled out an overall factor of $k$ from the action.  At leading order in $k$, this partition function is found by computing the  classical action of a Euclidean solution to the equations of motion.  The simplest such solution is just Euclidean AdS space, with periodically identified time coordinate. The contribution of this geometry to the partition function can be expanded in perturbation theory
\be
Z_{saddle}(\tau) \sim e^{-k S^{(0)}+S^{(1)}+k^{-1}S^{(2)}+\dots}
\label{Zsaddle}
\ee
The classical action $S^{(0)}$ is (see e.g. \cite{Maldacena:1998bw})
\be
e^{-k S^{(0)}} = |q|^{-2k}
\label{Scl}
\ee
where $q=e^{2\pi i \tau}$.

The one loop correction $S^{(1)}$ was derived indirectly in \cite{Maloney:2007ud,Yin:2007gv}, following the logic of Brown and Henneaux \cite{Brown:1986nw}.  Brown and Henneaux argued that the symmetry group relevant to general relativity with asymptotically AdS$_{3}$ boundary conditions is two copies of the Virasoro algebra.  This means that the partition function (\ref{Zsaddle}) must be the character of some representation of the Virasoro algebra:
\be
Z_{saddle}(\tau) = \Tr q^{L_{0}} \bar q^{\bar L_{0}}
\label{Ztrace}
\ee
Since the operators $L_{0}+\bar L_{0}$ and $L_{0}-\bar L_{0}$ are identified with energy and angular momentum operators, respectively, this is just the usual expression for a canonical ensemble partition function at fixed temperature and angular potential.
The classical action (\ref{Scl}) is interpreted as the contribution to (\ref{Ztrace}) of a ground state $|0\rangle$ of weight $L_{0} = \bar L_{0}=-k$.
The trace in equation (\ref{Ztrace}) is over the Hilbert space of perturbative excitations around this AdS$_{3}$ background, and the other states appearing in this trace will give the subleading corrections appearing in (\ref{Zsaddle}). These states are the Virasoro descendants of the ground state, found by acting on $|0\rangle$ with some combination of the Virasoro operators $L_{-n}$.
Including these states in the trace gives
\be
Z_{saddle}=|q|^{-2k}\prod_{n=2}^{\infty}{1\over |1-q^{n}|^{2}}
\label{Zexact}
\ee
The additional terms appearing in (\ref{Zexact}) are identified as $e^{S^{(1)}}$.\footnote{In fact, this expression must be one loop exact, because there is a unique representation of the Virasoro algebra with lowest weight. So there is no possible modification of the formula (\ref{Zexact}) -- aside from a renormalization of the coupling $k$ -- which is consistent with the Virasoro symmetry.}

In this paper we will compute the one-loop partition function (\ref{Zexact}) directly.  In particular, we will compute the one-loop determinant $\det \Delta^{(2)}$, where $\Delta^{(2)}$ is the kinetic operator for linearized graviton fluctuations around the background metric.  In computing the partition function, we must also include the Fadeev-Popov determinants arising due to gauge fixing.  These involve the determinants of a scalar Laplacian $\Delta^{(0)}$ and a vector field Laplacian $\Delta^{(1)}$.  Although the intermediate stages of this computation are quite complicated, the final answer takes a simple form:
\be
e^{-k S^{(0)}+S^{(1)}}= e^{-kS^{(0)}}{\det \Delta^{(1)}\over \sqrt{\det \Delta^{(0)}
\det\Delta^{(2)}}}= |q|^{-2k}\prod_{n=2}^{\infty} {1\over |1-q^{n}|^{2}}
\label{Sone}
\ee
exactly as was argued above using more indirect arguments.
This computation demonstrates directly that the structure of a conformal field theory emerges from quantum gravity in Anti-de Sitter space.  

We may apply the one-loop results derived in this paper to other locally hyperbolic geometries, in addition to thermal Anti-de Sitter space.  These geometries are quotients of the form $\bH_{3}/\Gamma$, where $\Gamma$ is a discrete group which acts freely on $\bH_{3}$.
These geometries may be thought of as higher genus generalizations of the BTZ black hole; they are Euclidean continuations of the ``wormhole'' solutions of 
\cite{Aminneborg:1997pz}.  The conformal boundary of one of these Euclidean geometries is a genus $g\ge2$ Riemann surface.  If one could compute exactly the partition function 
of quantum gravity on these backgrounds, this would determine uniquely the operator product expansion of the CFT dual to pure gravity in Anti-de Sitter space.  We are able to compute the one loop part of the action, which is a first step towards this ambitious goal.

Before discussing the technicalities of our computations, we will start by outlining our strategy, which utilizes the heat kernel approach to the computation of one-loop determinants in curved space-time.

\subsection{The Heat Kernel Method}

Our goal is to compute the partition function
$$
Z = \int D\phi \, e^{-g^{-2}S(\phi)}
$$
of a free quantum field $\phi$ to one loop in the coupling $g$.  Here $\phi$ denotes a scalar, vector or linearized metric field.
Since $\phi$ is a free field, the computation is -- in principle -- completely straightforward.
We start by writing the action as \footnote{We are neglecting total derivatives, which will not be important here.}
$$
S(\phi) = \int_{\cM} d^{3}x \,\sqrt{g}\, \phi\, \Delta \,\phi
$$
where $\Delta$ is a second order differential operator.  Of course, if $\phi$ is a gauge or graviton field then $\Delta$ will have a complicated tensor structure, which we suppress here.
As an operator on the space of normalizable functions on $\cM$, $\Delta$ will in general have both continuous and discrete spectrum of eigenvalues. For example, if $\cM$ is compact, then $\Delta$ has a discrete set of eigenvalues $\lambda_{n}$, in terms of which the one loop correction is
\be
S^{(1)}=-\frac{1}{2}\log \det(\Delta) = -\frac{1}{2}\sum_{n}\log \lambda_{n}\,.
\label{Sdet}
\ee
The manifolds $\cM$ considered in this paper are typically non-compact and homogeneous, on which $\Delta$ also has a continuous spectrum. This gives a divergent contribution to (\ref{Sdet}) proportional to the volume of $\cM$.
This divergence can be absorbed into a local counterterm, which describes the renormalization of the cosmological constant (or alternatively of Newton's constant) at one-loop.

In practice, the computation of $S^{(1)}$ is quite difficult, especially for gauge and graviton fields.  The most straightforward procedure is to find a complete basis of normalizable eigenfunctions $\{\psi_{n}\}$ obeying $\Delta \psi_{n} = \lambda_{n} \psi_{n}$ and compute the sum (\ref{Sdet}) directly.  This is a formidable task.  Instead, we will use a heat kernel approach.  The heat kernel
$K(t,x,y)$ is a function of two points $x$ and $y$ on $\cM$, as well as an auxiliary ``time'' variable $t$.  It can be defined as
\be
K(t,x,y) = \sum_{n}e^{-\lambda_{n} t}\psi_{n}(x) \psi_{n}(y)
\label{Kdef1}
\ee
where, as above, we are suppressing the complicated tensor structure.
In writing this we have normalized the eigenfunctions $\psi_{n}$ so that
$$
\sum_{n} \psi_{n}(x) \psi_{n}(y) = \delta^{d}(x,y),~~~~~\int_{\cM} d^{3}x\, \sqrt{g}\, \psi_{n}(x) \psi_{m}(x) = \delta_{nm}
$$
The trace of the heat kernel
$$
\int_{\cM} d^{3}x \, \sqrt{g}\,K(t,x,x) = \sum_{n} e^{-\lambda_{n}t}
$$
is a function of $t$ which encodes information about the spectrum of $\Delta$.  For example, we may compute
$$
S^{(1)} = -\frac{1}{2}\sum_{n}\log \lambda_{n} = \frac{1}{2}\int^{\infty}_{0^{+}}{dt\over t} \int_{\cM} d^{3}x \sqrt{g}\, K(t,x,x)\,.
\label{detform}
$$

The advantage of the heat kernel method is that $K(t,x,y)$ satisfies
the differential equation
\be
(\partial_{t} + \Delta_{x}) K(t,x,y)=0
\label{Kdef2}
\ee
with boundary conditions at $t=0$
\be
K(0,x,y) = \delta(x,y)
\label{Kbc}
\ee
where $\delta(x,y)$ is the appropriate delta function on $\cM$.
In practice, we may define the heat kernel as the unique solution to the differential equation (\ref{Kdef2}) with boundary conditions (\ref{Kbc}).  This definition is more easily computable than the original definition (\ref{Kdef1}).
In particular, since hyperbolic space $\bH_{3}$ is a symmetric space, it is possible to solve this differential equation, including the complicated tensor structure.
This is the computation described in section 2.  The resulting one-loop determinant (\ref{detform}) is computed in section 3.

The heat kernel method is particularly well adapted to the computation of the one loop partition function on quotient spaces $\cM=\bH_{3}/\Gamma$.  Let us start by imagining that we have found the heat kernel $K^{\bH_{3}}(t,x,y)$ on $\bH_{3}$.  The differential equation (\ref{Kdef2}) is linear, so we may find the heat kernel on $\bH_{3}/\Gamma$ using the method of images
\be
K^{\bH_{3}/\Gamma}(t,x,y) = \sum_{\gamma\in\Gamma}K(t,x,\gamma y)
\label{Kimages}
\ee
For example, thermal AdS and the BTZ black hole are both quotients of AdS by $\bZ$.  The Euclidean space $\cM = \bH_{3}/\bZ$ turns out to be a solid torus endowed with a metric of constant negative curvature.  In this case the sum (\ref{Kimages}) is a relatively simple sum over $\bZ$ which can be computed.  This computation is described in section 4.

For other choices of the group $\Gamma$, the Euclidean space $\bH_{3}/\Gamma$ is more complicated.  We are interested in cases where the group $\Gamma$ acts freely on $\bH_{3}$.  In this case $\bH_{3}/\Gamma$ is a smooth manifold of constant negative curvature, and the conformal boundary of $\bH_{3}/\Gamma$ is a Riemann surface of genus $g\ge 2$.  For some choices of $\Gamma$, $\bH_{3}/\Gamma$ is a solid handlebody of genus $g\ge 2$ endowed with a metric of constant negative curvature.  For other choices, the geometry of $H_{3}/\Gamma$ is more complicated.
There is a rich mathematical theory -- that of the Selberg trace formula and its generalizations -- where the sum over elements $\gamma \in \Gamma$ is used to compute the spectrum of differential operators on $\bH_{d}/\Gamma$.  For scalar and vector fields, our computations precisely reproduce the results of the Selberg trace formula (see e.g. \cite{Bytsenko:1997sr} and references therein).  To our knowledge, the Selberg trace formula has not been successfully generalized to the graviton case.\footnote{This was attempted for AdS$_{3}$ in \cite{Bytsenko}, whose result differs from ours.  We believe this is due to an error in the computation of \cite{Bytsenko}.}  Our computation may therefore be viewed as a brute force derivation of the Selberg trace formula in this context.  This computation is described in section 5.

\section{Heat Kernels in Hyperbolic Space $\bH_{3}$}

In this section we compute the heat kernel for scalar, gauge and graviton fields in Euclidean Anti-de Sitter space $\bH_{3}$.

\subsection{Scalar fields}
\label{scalars}

We consider a scalar field $\phi$ of mass $m$ in hyperbolic 3-space $\bH_{3}$. The action is
\begin{eqnarray}
S &&= {1\over 2}\int d^{3}x \sqrt{g} \left(\partial_{\mu}\phi\partial^{\mu}\phi + m^{2}\phi^{2}\right) \\
&&= {1\over 2} \int d^{3}x \sqrt{g} \phi \left(-\nabla^{2}+m^{2}\right)\phi
\end{eqnarray}
where
$\nabla^2=\nabla^{\mu} \nabla_{\mu}$ is the scalar Laplacian on $\bH_{3}$.
In the second line we have discarded a total derivative term, which vanishes provided $\phi$ obeys suitable boundary conditions at the boundary of $\bH_{3}$.

The heat-kernel $K(t,x,x')$ is a solution of the differential equation
\begin{equation}
(\nabla^2_{x}-m^2) K(t,x,x')= \partial_t K(t,x,x')
\label{heat-equation}
\end{equation}
where $x,x'$ are coordinates on $\bH_{3}$ and $\nabla^{2}_{x}$ is the Laplacian acting on $x$.
The boundary condition on $K(t,x,x')$ at $t=0$ is
\begin{equation}
K(0,x,x')=\delta^3(x,x')\,,
\label{heat-equationn}
\end{equation}
where $\delta^3(x,x')={1\over \sqrt{g(x)}}\delta^3(x-x') $.

 We will use the following metric on $\bH_{3}$
\begin{equation}
ds^2 = g_{\mu\nu} dx^{\mu} dx^{\nu} = {dy^2 + dz d\bar z\over y^2}\,,
\label{H3-metric}
\end{equation}
where $y>0$ and $z$ is a complex coordinate.  Since the space $\bH^{3}$ is maximally
symmetric, the heat kernel $K(t,x,x')$ will depend on $x$ and $x'$ only through the geodesic distance
\begin{equation}
r(x,x') \equiv \mbox{arccosh} (1+u(x,x'))\,,
\end{equation}
where we have defined the ``chordal distance''
\begin{equation}
u(x,x') \equiv {(y-y')^2+|z-z'|^2\over 2 yy'}\,.
\label{chordal}
\end{equation}
Acting on a function of the geodesic distance $r(x,x')$, the scalar Laplacian is
\begin{equation}
\nabla^2= u(u+2) \partial^2_u + 3 (u+1) \partial_u = \partial^2_r
+ 2 \coth r \,\partial_r.
\end{equation}

It is straightforward to solve the differential equation (\ref{heat-equation}) subject to the boundary condition (\ref{heat-equationn}).
The answer for the scalar heat kernel on $\bH_{3}$ is
\begin{equation}
K^{{\mathbb H}_3}(t,r) = \frac{e^{-(m^2+1)t-\frac{r^2}{4 t}}}{(4\pi t)^{3/2}} \,
\frac{r}{\sinh r}\,.
\label{scalar-heat}
\end{equation}
This was first described in reference \cite{Camporesi:1990wm} (see also \cite{Mann:1996ze}).

\subsection{Vector fields}
\label{vectors}

We now consider a $U(1)$ gauge field $A_{\mu}$ on $\bH_{3}$, with action
\begin{equation}
S = {1\over 4} \int d^{3}x \sqrt{g}  F_{\mu\nu}F^{\mu\nu}
\end{equation}
In order to study the heat kernel, it is necessary to fix a gauge.
We will work in Feynman gauge, where the gauge fixed action is
\begin{eqnarray}
S &&= \int d^{3}x \sqrt{g} \left({1\over 4} F_{\mu\nu}F^{\mu\nu} + {1\over 2} \left(\nabla_{\mu}A^{\mu} \right)^2\right) \\
 &&= {1\over 2} \int d^{3}x \sqrt{g} A_{\mu} \left(-g^{\mu\nu}\nabla^{2} + R^{\mu\nu}\right)A_{\nu}
 \label{gaugeaction}
\end{eqnarray}
As in the scalar case, in the second line we have discarded a total derivative term.

This gauge fixing procedure introduces Fadeev-Popov ghosts $b$ and $c$, which are anti-commuting scalar fields.  Their action is
\begin{equation}
S_{ghost} = - \int d^{3}x \sqrt{g}~ b \nabla^{2}c
\label{gaugeghost}
\end{equation}
The heat kernel for these ghost fields can be computed exactly as in Section \ref{scalars}.

The heat-kernel for a $U(1)$ gauge field
on $\bH_3$ is a bitensor $K_{\mu \nu'}(t,x,x')$ which solves the differential equation
\begin{equation}
\begin{aligned}
&{\Delta_\mu}^{\nu}
K_{\nu \nu'}(t,x,x') = -\partial_t K_{\mu\nu'}(t,x,x')
\label{gauge-heat-eq}
\end{aligned}
\end{equation}
where $\Delta^{\nu}_{\mu}$ is the  gauge field kinetic operator appearing in the action (\ref{gaugeaction})
\begin{equation}
{\Delta_\mu}^{\nu} = -{\delta_\mu}^\nu \nabla^2 + {R_\mu}^\nu = -(\nabla^2 + 2) {\delta_\mu}^{\nu}\,.
\label{vectordiff}
\end{equation}
The boundary condition at $t=0$ is
\begin{equation}
K_{\mu \nu'}(0,x,x') = g_{\mu\nu'}(x) \delta^3(x,x')\,.
\label{gauge-heat-eqq}
\end{equation}

As in the scalar case, since $\bH_{3}$ is a maximally symmetric space the heat kernel is a function only of the geodesic distance $r(x,x')$, or alternatively of the chordal distance $u(x,x')$ defined in (\ref{chordal}).
This implies that $K_{\mu\nu'}$ can be written as a linear combination of $\p_{\mu}u\p_{\mu'}u$ and $\p_{\mu}\p_{\mu'}u$, which form a basis for the space of $(1,1)$ bitensors constructed out of $u(x,x')$.  Appendix A lists a few useful identities involving $u(x,x')$ and the various tensors constructed from $u(x,x')$ (see also \cite{Freedman}).  Using the properties listed in Appendix A it follows that the heat kernel can be written as
\begin{equation}
K_{\mu \nu'}^{}(t,x,x') =  F(t,u) \partial_{\mu}
\partial_{\nu'}u +
\partial_{\mu} \partial_{\nu'} S(t,u)\,,
\label{heataa}
\end{equation}
and that
\begin{equation}
\left(\nabla^2 +2\right)K^{}_{\mu\nu'}=\left(\nabla^2 F +
F\right)
\partial_{\mu}\partial_{\nu'} u + \partial_{\mu}\partial_{\nu'} \left(\nabla^2
S - 2 \int_u^\infty F(t,v) dv\right)\,.
\label{gaugeheat}
\end{equation}
The heat equation (\ref{gauge-heat-eq}) becomes
\begin{equation}
\begin{aligned}
&\left(\nabla^2 + 1 \right) F(t,u) = \partial_t F(t,u) \\
&\nabla^2 S(t,u) - 2 \int_u^\infty F(t,v) dv= \partial_t
S(t,u)\,.
\end{aligned}
\end{equation}
and the boundary condition (\ref{gauge-heat-eqq}) becomes
\begin{equation}
\begin{aligned}
&F(0,u)=-\delta^3(x,x')\,, \\
&\partial_u S(0,u)=u\partial^2_u S(0,u)=0\,. \label{bcond}
\end{aligned}
\end{equation}
The second condition ensures that the terms in (\ref{heataa})
involving derivatives of $S$ will not contain delta functions
at $t=0$. \footnote{More precisely,
$\partial_u S$ and $\partial_u^2 S$ are
delta functions in $r$ at $t=0$ which integrate to zero when multiplied by the
measure $r^2 dr$, so can be ignored.}
The solution is \footnote{The equation for $S$ is solved by Fourier transforming the equation for
$\sinh r S(r)$ with respect to $r$.}
\begin{equation}
\begin{aligned}
&F(t,r)=-\frac{e^{-\frac{r^2}{4 t}}}{(4\pi t)^{3/2}} \,
\frac{r}{\sinh r}\,, \\
&S(t,r)=\frac{4}{(4\pi)^{3/2}}\frac{e^{-\frac{r^2}{4t}}}{\sinh
r}\sqrt{t}
 \int_0^1 d\xi\, e^{-t(1-\xi)^2} \sinh{r \xi}\,.
\label{Ssol}
\end{aligned}
\end{equation}

Although we have just considered massless vector fields, it is straightforward to generalize these results to find the heat kernel of a massive vector field.  The differential operator 
$\nabla^{2}+2$ appearing in (\ref{vectordiff}) is replaced by $\nabla^{2}+2-m^{2}$.  The heat kernel takes exactly the same form as in (\ref{Ssol}), except multiplied by a factor of $e^{-m^{2}t}$; this is the same exponential factor appearing in the massive scalar heat kernel (\ref{scalar-heat}). 

\subsection{Graviton}
\label{gravity}

We will now consider a linearized graviton perturbation $h_{\mu\nu}$ around an $\bH_{3}$ background.
The Einstein-Hilbert action for gravity with a negative cosmological constant is
\begin{equation}
S_{GR} = -{1\over 16\pi G} \int d^3x \sqrt{g} (R+2)\,.
\label{Sgr}
\end{equation}
The form of the kinetic term for the metric perturbation $h_{\mu\nu}$ will depend on the choice of gauge.
We will use the gauge of \cite{Christensen:1979iy}, where we add to (\ref{Sgr}) the gauge fixing term
\begin{equation}
S_{GF} = {1\over 32\pi G} \int d^3x \sqrt{g}\, \nabla^\mu \!\left(h_{\mu\sigma}-{1\over 2}g_{\mu\sigma}h\right)
\nabla^\nu \!\left({h_\nu}^\sigma-{1\over 2}{\delta_\nu}^\sigma h\right).
\end{equation}
It is convenient to separate out the traceless and pure trace parts of $h_{\mu\nu}$
\begin{equation}
\phi_{\mu\nu} = h_{\mu\nu} - {1\over 3} g_{\mu\nu} {h^\rho}_\rho,~~~~~~ \phi=h^{\rho}_{\rho}
\end{equation}
The gauge fixed action is
\begin{equation}\label{phigravv}
S_{GR}+S_{GF}= -{1\over 32 \pi G}\int d^{3}x \sqrt{g}\left\{ \frac{1}{2}\phi_{\mu\nu}\left(g^{\mu\rho}g^{\nu\sigma}\nabla^{2}+2R^{\mu\rho\nu\sigma}\right)\phi_{\rho\sigma} -{1\over 12} \phi\left(\nabla^{2}-4\right)\phi\right\}
\end{equation}
Note that the kinetic term for the trace mode $\phi$ has the wrong sign.  To deal with this mode, we will use the standard procedure of \cite{Gibbons:1978ac}: we Wick rotate $\phi\to i\phi$, so that the kinetic term becomes positive definite.  Then $\phi$
is just a scalar field with mass $m^{2}=4$; the heat kernel for such a field was described in Section \ref{scalars}.

As above, the gauge-fixing procedure requires us to introduce a Fadeev-Popov ghost field, which in this case is a complex valued vector $\eta_\mu$.  The action is
\begin{equation}\label{ghostact}
S_{ghost}=\frac{1}{32\pi G}\int d^{3}x \sqrt{g} \, {\bar \eta}_{\mu}\left(-g^{\mu\nu}\nabla^{2}-R^{\mu\nu}\right) \eta_{\nu}
\end{equation}
The heat kernel for this ghost field is that of a vector field with mass $m^2=4$, which is $e^{-4t}$ times the
heat kernel of a massless vector field in Feynman gauge described in Section \ref{vectors}.

We will now compute the heat kernel $K_{\mu\nu,\mu'\nu'}(t,x,x')$ for $\phi_{\mu\nu}$.  This heat kernel equation is
\begin{equation}
\Delta^{\rho\sigma}_{\mu\nu}K_{\rho\sigma,\mu'\nu'}(t,x,x')=\partial_t K_{\mu\nu,\mu'\nu'}(t,x,x')
\label{grav_heat_eq}
\end{equation}
where $\Delta^{\rho\sigma}_{\mu\nu}$ is the differential operator appearing in (\ref{phigravv})
\begin{equation}
\Delta^{\rho\sigma}_{\mu\nu}= \delta^{\rho}_{\mu}\delta^{\sigma}_{\nu}\nabla^{2}+ 2R_{\mu~\nu}^{~\rho~\sigma}= \delta^{\rho}_{\mu}\delta^{\sigma}_{\nu}\left(\nabla^{2}+2\right)
\label{gravdiff}
\end{equation}
The boundary condition at $t=0$ is
\begin{equation}
K_{\mu\nu,\mu'\nu'}(0,x,x') = {1\over 2}\left(g_{\mu\mu'} g_{\nu\nu'}+g_{\mu\nu'} g_{\nu\mu'}
- {2\over 3} g_{\mu\nu} g_{\mu'\nu'}\right)\delta^3(x,x')
\label{grav_heat_eqq}
\end{equation}
Finally, since $\phi_{\mu\nu}$ is traceless, the heat kernel will be traceless as well:
\begin{equation}
g^{\mu\nu}K_{\mu\nu,\mu'\nu'}(t,x,x') = g^{\mu'\nu'} K_{\mu\nu,\mu'\nu'}(t,x,x') = 0
\label{grav_heat_eqqq}
\end{equation}

As in the scalar and vector cases, the heat kernel on ${\mathbb H}_3$ must be a function of the chordal distance $u(x,x')$.  There are six (2,2) bitensors which can be written as a function of $u(x,x')$; they are described in Appendix A (and also \cite{Freedman}).  From this fact, along with the symmetry of $K_{\mu\nu.\mu'\nu'}$ under $x\to x'$, it follows that the heat kernel may be expressed as a linear combination of the following five terms
\begin{eqnarray} \label{tensorstr}
K_{\mu\nu,\mu'\nu'}^{{\mathbb H}_3}(t,x,x') &=& (\partial_\mu\partial_{\mu'} u \partial_\nu\partial_{\nu'}u
+\partial_\mu\partial_{\nu'} u \partial_\nu\partial_{\mu'}u) G(t,u) + g_{\mu\nu} g_{\mu'\nu'} H(t,u)\nonumber\\
&& + \nabla_{(\mu}\! \left[ \partial_{\nu)} \partial_{(\mu'} u \partial_{\nu')} u X(t,u) \right]
+ \nabla_{(\mu'}\! \left[ \partial_{\nu')} \partial_{(\mu} u \partial_{\nu)} u X(t,u) \right] \nonumber\\
&& + \nabla_{(\mu}\! \left[ \partial_{\nu)}u \partial_{\mu'} u \partial_{\nu'} u Y(t,u) \right]
+ \nabla_{(\mu'}\! \left[ \partial_{\nu')}u \partial_{\mu} u \partial_{\nu} u Y(t,u) \right] \nonumber\\
&& +\nabla_\mu \left[ \partial_\nu u Z(t,u) \right] g_{\mu'\nu'} + \nabla_{\mu'} \left[ \partial_{\nu'} u Z(t,u) \right] g_{\mu\nu}.
\label{kexpansion}
\end{eqnarray}
where $G$, $H$, $X$, $Y$, and $Z$ are five functions of $t$ and $u(x,x')$.

In terms of these functions the heat equation
(\ref{grav_heat_eq}) becomes
\begin{eqnarray}
&&\nabla^2 G = \partial_t G,\nonumber\\
&& \nabla^2 H - 4H-4G - 8(u+1) \int_u^\infty G(t,v) dv = \partial_t H, \nonumber\\
&& \nabla^2 X + 2(u+1) \partial_u X+4 X + 4 (u+1) Y + 4 G = \partial_t X,\\
&& \nabla^2 Y + 6 (u+1) \partial_u Y + 2 \partial_u X + 7 Y = \partial_t Y,\nonumber\\
&& \nabla^2 Z + 2(u+1) \partial_u Z - Z + 2 Y + 4 \int_u^\infty G(t,v) dv = \partial_t Z.\nonumber
\label{gravitoneom}
\end{eqnarray}
The boundary conditions (\ref{grav_heat_eqq}) are
\begin{equation}
\begin{aligned}
& G(0,u) = {1\over 2}\delta^3(x,x')\,,\\
& H(0,u) = -{1\over 3}\delta^3(x,x')\,,\\
& X(0,u)=u\partial_u X(0,u)=0\,,\\
& uY(0,u)=u^2\partial_u Y(0,u)=0\,,\\
& Z(0,u)=u\partial_u Z(0,u)=0\,.\\
\end{aligned}
\end{equation}
The traceless condition (\ref{grav_heat_eqqq}) is
\begin{eqnarray}
&& 2 G + 5 X + 2 (u+1) \partial_u X + 2u(u+2) \partial_u Y + 7(u+1) Y+3 \partial_u Z = 0,\label{traceless}\\
&& 2 G + 3H + (u^2+2u+2) X + u(u+1)(u+2) Y + 6(u+1) Z + u(u+2) \partial_u Z =0.\nonumber
\end{eqnarray}

The solution of these equations is somewhat involved.  The first two equations in (\ref{grav_heat_eq}) are solved by
\begin{eqnarray}
&& G(t,u) = {e^{-{r^2\over 4t}-t}\over 2(4\pi t)^{3\over 2}} {r\over\sinh r},\nonumber\\
&& H(t,u) = {e^{-{r^2\over 4 t} -5 t}\over 2(4\pi t)^{3\over 2}} \left({1\over 3}-e^{4 t}\right) {r\over \sinh r}
-{e^{-{r^2\over 4 t}-t}\sqrt{t}\over 2 \pi^{3\over 2}\sinh r} \int_0^1 d\xi e^{-4 t(1-\xi)^2} \sinh(2 r \xi)~~~~~~~~
\end{eqnarray}
To solve for $X$ and $Y$, it is useful to define the function $V(t,u)$ by
\begin{equation}\partial_u V(t,u) = X+ 2\int_u^\infty Y(t,v) dv\end{equation}The heat kernel equations (\ref{grav_heat_eq}) imply that
\begin{equation}
\nabla^2 V - 3 V - 4 \int_u^\infty G(t,v) dv = \partial_t V.
\end{equation}
which is solved by
\begin{equation}
V(t,u) = -{e^{-{r^2\over 4 t}}\sqrt{t}\over 2\pi^{3\over 2}\sinh r} \int_0^1 d\xi e^{-t(2-\xi)^2} \sinh(r \xi).\nonumber
\end{equation}
The solution for $Y(t,u)$ is simplest to describe in terms of its triple integral in $u$.  If we define
$\tilde Y(t,u)$ by
$Y(t,u) = \partial_u^3 \tilde Y(t,u)$, then
\begin{equation}
\begin{aligned}
&\tilde Y(t,u) = {e^{-2t}\sqrt{t}\over 2(4\pi)^{3\over 2} \sinh r} \int_0^1 d\xi e^{-{(r-2t\xi)^2\over 4t}}
\left[ e^t \sinh(2 t\xi)-2 (e^{-3t(1-\xi)}+e^{t-2r}\sinh(t \xi))\sinh(t \xi)\right.
\\
&~~~~~~\left.
+ 2 (e^{-2t(1-\xi)}-1) \cosh r \sinh(2 t\xi)+2 \sinh r \left(\cosh(2t)-\cosh(2t\xi)\right)\right]
\end{aligned}
\end{equation}
One can then recover $X$ from $X(t,u) = 2 \partial_u^2\tilde Y(t,u) + \partial_u V(t,u)$.
Finally, $\partial_u Z(t,u)$ can be determined from the first line of (\ref{traceless}).  The expressions for $X$ and $Z$ are rather lengthy, so we will not write them here.

As in the previous section, it is reasonably straightforward to generalize these results to find the heat kernel of a massive spin two field.  The differential operator 
$\nabla^{2}+2$ appearing in (\ref{gravdiff}) is replaced by $\nabla^{2}+2-m^{2}$.  This means that the solution for the heat kernel described above is simply multiplied by a factor of $e^{-m^{2}t}$.

\section{One-loop determinants on $\bH_{3}$}

In the previous section we computed the heat kernels for scalar, gauge and graviton excitations in Euclidean Anti-de Sitter space $\bH_{3}$.  In this section we will use these results to compute the corresponding one loop determinants in $\bH_{3}$.  These one loop determinants simply multiply the partition function by an overall constant; they describe the renormalization of the cosmological constant $\Lambda$ (or alternatively of Newton's constant $G_{N}$) at one-loop.  Since these determinants can be absorbed into a local counterterm, they do not provide any new physical information.  The reader who is not interested in the computational details may therefore skip this section.

Nevertheless, the computations of this section are useful for two reasons.  First, they are illustrative of the basic technique, which will be applied in more physically interesting settings in the next section.  Second, these one-loop determinants were computed for arbitrary spin fields in \cite{Camporesi:1994ga} using a different technique, so this section provides a useful check of the heat kernels computed above.

\subsection{Scalar Field}

The one-loop contribution of a scalar field of mass $m$ to the effective action is
\begin{equation}
S^{(1)} = -\frac{1}{2}\log \det \left(-\nabla^{2}+m^{2}\right) = \frac{1}{2}\int_{0}^{\infty} {dt\over t} \int d^{3}x \sqrt{g} K^{\bH_{3}} (t,x,x)
\end{equation}
where $K^{\bH_{3}}(t,x,x')$ was computed in (\ref{scalar-heat}).  Setting $r=0$ in (\ref{scalar-heat}) and performing the integral over $x$ gives
\begin{equation}
S^{(1)} = \frac{1}{2} Vol(\bH_{3}) \int {dt\over t} {e^{-(m^{2}+1)t}\over (4\pi t)^{3/2}
}
\end{equation}
There are two divergences appearing in this integral.  The first is an infrared divergence coming from the integral over $\bH_{3}$ which gives a (divergent) factor of $Vol(\bH_{3})$.  The second is an ultraviolet divergence coming from $t\to0$ behavior of the integrand.  Of course, these are both removed by a local counterterm.  It is worthwhile noting that the $t$ integral can be defined by analytic continuation -- it is just a Gamma function with a negative argument -- allowing us to remove the ultraviolet divergence.  The answer is
\begin{equation} \label{scalaroneloop}
S^{(1)}=Vol(\bH_{3}) {(m^{2}+1)^{3/2} \over 12\pi}
\end{equation}
This result agrees with that derived in \cite{Camporesi:1994ga}.

\subsection{Vector Field}

The trace of the heat kernel for a massless vector field 
is
\begin{equation}\label{withtrans}
\begin{aligned}
{\rm Tr} K_{vec}^{\bH_3}(t) & = g^{\mu\nu'} K_{\mu\nu'}^{\bH_3}(t,x,x) \\
& = -3\left.\left[F(t,u)+\partial_u S(t,u)\right]\right|_{u=0} \\
&= {e^{-t}+ 2+4t\over (4\pi t)^{3\over 2}}\,,
\end{aligned}
\end{equation}
where $K_{\mu\nu'}^{\bH_3}(t,x,x')$ was computed in (\ref{heataa}).
This result includes a contribution from the longitudinal mode of the vector field.

To obtain the heat kernel just for the transverse components of a vector field, we must subtract from (\ref{withtrans}) the heat kernel of a scalar field.
For a transverse vector field of mass $m$, the resulting one loop determinant is \begin{equation}
\begin{aligned}
-\log \mbox{det}_{(1)} \Delta &= Vol(\bH_{3})
\int^{\infty}_0 \frac{dt}{t}e^{-m^2 t}\left({\rm Tr}\, K^{\bH_3}_{vec}-\frac{e^{-t}}{(4\pi t)^{3/2}} \right)\\
&=Vol(\bH_{3})\int^{\infty}_0 \frac{dt}{t} \frac{e^{-m^2t}}{(4\pi t)^{3/2}}(2+4t)=Vol(\bH_{3})\frac{m^3-3m}{3\pi}\,,
\label{transvec}
\end{aligned}
\end{equation}
As in the previous section, we have regularized the UV divergence by analytically continuing the $t$-integral. This result (\ref{transvec}) agrees precisely with the one obtained in \cite{Camporesi:1994ga}. 

We should emphasize that (\ref{transvec}) vanishes when $m=0$. Thus the contribution to the one-loop effective action for a $U(1)$ gauge field, including the (complex) ghost, is just minus that of a massless scalar field in (\ref{scalaroneloop}). 


\subsection{Graviton}

The trace of the heat kernel for the symmetric traceless tensor mode $\phi_{\mu\nu}$ is
\begin{equation}
\begin{aligned}
{\rm Tr}\, K^{\bH_3}_{\phi_{\mu\nu}}&=g^{\mu\mu'}g^{\nu\nu'} K^{\bH_3}_{\mu\nu,\mu'\nu'}(t,x,x)\\
&=\left.\left[12(G+X)+3H+6(u+1)Z\right]\right|_{u=0}\\ &=\frac{e^{-5t}}{(4\pi t)^{3/2}} \left(1+2e^t(1+2t)+2e^{4t}(1+8t)\right)\,
\label{qwerty}
\end{aligned}
\end{equation}
where $K^{\bH_3}_{\mu\nu,\mu'\nu'}(t,x,x')$ was computed in (\ref{tensorstr}).  This heat kernel includes contributions from longitudinal polarizations of $\phi_{\mu\nu}$.

The one-loop determinant for a massless transverse, symmetric 2-tensor is obtained by subtracting from (\ref{qwerty}) the contribution of a massive vector
with mass squared equal to 4. The one-loop determinant for a massive
transverse symmetric 2-tensor is found by multiplying this heat kernel by a factor of $e^{-m^{2}t}$, as described in section 3.  Putting this all together, we find that the one-loop determinant for a massive, transverse symmetric 2-tensor is:
\begin{equation}\label{tensgam}
\begin{aligned}
-\log \mbox{det}_{(2)} \Delta &= Vol(\bH_{3})\int^{\infty}_0 \frac{dt}{t}e^{-m^2 t}
\left({\rm Tr}\, K^{\bH_3}_{\phi_{\mu\nu}}-e^{-4 t}{\rm Tr}\, K^{\bH_3}_{vec} \right)\\
&=Vol(\bH_{3})\int^{\infty}_0 \frac{dt}{t} \frac{e^{-(m^2+1)t}}{(4\pi t)^{3/2}}(2+16t)=\frac{Vol(\bH_{3})}{3\pi}(1+m^2)^{1/2}\left(m^2-11\right)\,,
\end{aligned}
\end{equation}
which again matches the result of \cite{Camporesi:1994ga}. 

The one-loop determinant for a linearized graviton fluctuation is found by adding to (\ref{qwerty}) the contributions from 
the complex massive vector ghost $\eta_\mu$ and the trace mode $\phi$.  The result is
\begin{equation}\label{gravgam}
\begin{aligned}
S^{(1)}_{grav} &= \frac{1}{2} Vol(\bH_{3})\int^{\infty}_0 \frac{dt}{t}
\left({\rm Tr}\, K^{\bH_3}_{\phi_{\mu\nu}}-2 e^{-4 t}{\rm Tr}\, K^{\bH_3}_{vec}+e^{-4t}{\rm Tr}\,K^{\bH_3} \right)\\
&=\frac{1}{2}Vol(\bH_{3})\int^{\infty}_0 \frac{dt}{t} \frac{1}{(4\pi t)^{3/2}}\left[2e^{-t}(1+8t)-2e^{-4t}(1+2t) \right]\\
&=-\frac{13}{6\pi}Vol(\bH_{3}).
\end{aligned}
\end{equation}
Once again, we have regularized the UV divergence by analytically continuing the $t$-integral.

\section{One-loop Determinants in Thermal AdS}

In section 2 we described the computation of the heat kernel on $\bH_{3}$ for scalar, gauge and graviton fields.  In this section we use these results to compute the one loop determinants on the quotient $\bH_{3}/\bZ$.  As described in section 1, this can be done using the method of images. We start by recalling briefly the geometry of thermal Anti-de Sitter space.

\subsection{The Geometry of $\bH_{3}/\bZ$}

Our goal is to study the one-loop effective action of quantum field theory in Anti-de Sitter space-time at finite temperature $\beta$ and angular potential $\theta$.  According to the usual rules of quantum field theory, we periodically identify \begin{equation}(t,\phi)\sim(t+\beta,\phi+\theta) \label{idents}\end{equation} where $t$ is the Euclidean time coordinate and $\phi$ an angular coordinate.  This means that the canonical ensemble partition function for a given $(\beta,\theta)$ is equal to the Euclidean partition function on $\bH_{3}/\Z$ where the group $\bZ$ is the group generated by the identification (\ref{idents}).

To describe the geometry of $\bH_{3}/\Z$, it is helpful to combine the two parameters $(\beta, \theta)$ into a single complex quantity $\tau= \frac{1}{2\pi}(\theta + i\beta)$.  In terms of the metric (\ref{H3-metric}) on $\bH_{3}$, $\Z$ is generated by an element $\gamma$ of the isometry group $SL(2,\bC)$ of $\bH_{3}$:
\begin{equation}
\gamma (y,z) \to (|q|^{-1} y\,,q^{-1} z)\,,
\label{gamma-act}
\end{equation}
where $q=e^{2\pi i\tau}$.  Geometrically, $\bH_{3}/\bZ$ can be thought of as a solid torus endowed with a metric of constant negative curvature, in the same way the $\bH_{3}$ is viewed as a unit ball with a metric of constant negative curvature.  The parameter  $\tau$ is the modulus of the  $T^{2}$ boundary of $\bH_{3}/\bZ$.

Later in this section we will find it useful to use a different set of coordinates on $\bH_{3}/\bZ$.  We define the the polar coordinates $(\rho, \theta, \phi)$ by
\begin{equation}
y = \rho \sin\theta\,,~~~~~z = \rho \cos\theta e^{i \phi}\end{equation}On ${\mathbb H}_3/\Z$ these coordinates run from $1\le\rho<e^{2\pi \tau_2}$, $0\le\theta<\pi/2$ and $0\le\phi<2\pi$.

We should emphasize that $\bH_{3}/\bZ$ is also the geometry of the Euclidean BTZ black hole.  In particular, if we let $\frac{1}{2\pi}(\theta'+i\beta')=-1/\tau$, the geometry described above is the Euclidean BTZ black hole with inverse Hawking temperature $\beta'$ and the angular potential $i\theta'$.  So although we have used so far the language of thermal field theory, we are at the same time computing the free energy of the BTZ black hole at one-loop.  This free energy determines the thermodynamic properties of the black hole.  So the computation of the one loop partition function gives, in particular, the one-loop correction to the BTZ entropy.  This was discussed in more detail in \cite{Maloney:2007ud}.

\subsection{Scalar fields}\label{solidtorus}

The heat-kernel on ${\mathbb H}_3/\Z$ can be obtained from the one on
${\mathbb H}_3$ given in (\ref{scalar-heat}) by the method of images:
\begin{equation}
K^{{\mathbb H}_3/\Z}(t,x,x') = \sum_{n \in \mathbb{Z}} K^{{\mathbb H}_3}\left(t,r(x,\gamma^n
x')\right)\,.
\end{equation}
It follows that the scalar one-loop determinant on ${\mathbb H}_3/\Z$ is
\begin{equation}
\begin{aligned}
-\log \det \Delta &=\int_0^{\infty} \frac{dt}{t} \int d^3x \sqrt{g}
K^{{\mathbb H}_3/\Z}
(t,x,x) \\
&= \mbox{vol}({\mathbb H}_3 / \Z) \int_0^{\infty} \frac{dt}{t}
\frac{e^{-(m^2+1)t}}{(4\pi t)^{3/2}} + \sum_{n \neq 0} \int_0^{\infty}
\frac{dt}{t} \int_{{\mathbb H}_3 / \Z} \!\!\!\!\! d^3 x \sqrt{g}
K^{{\mathbb H}_3}\left(t,r(x,\gamma^n x)\right)\,. \label{scalar-logdet}
\end{aligned}
\end{equation}
The first term was discussed in the previous section; it is divergent, and proportional to $\mbox{vol}({\mathbb H}_3 / \Z)$.  This term describes the renormalization of the cosmological constant at one loop, and can be canceled by a local counterterm.

The other terms in (\ref{scalar-logdet}) are more interesting. For a given $n\not=0$, it is convenient to replace the angular $\theta$ coordinate by the following $r$ coordinate
\begin{equation}
r\equiv r(x,\gamma^n x) = \mbox{arccosh} \left(1+ 2 \sinh^2 \pi n \tau_2 +
2 |\sin \pi n \tau|^2 \cot^2\theta \right)
\end{equation}
which lives in the interval $r\in [2\pi n\tau_2,\infty)$.
The measure is
\begin{equation}
d^3 x \sqrt{g} = \frac{d\rho}{\rho}\, d\phi\, d\theta
\frac{\cos\theta}{\sin^3 \theta} = \frac{d\rho}{\rho}\, d\phi\, \frac{dr \sinh r}{4
|\sin \pi n \tau|^2}\,.\label{change}
\end{equation}
The integral over $r$ in (\ref{scalar-logdet}) is
\begin{equation}
\int_{2\pi n\tau_2}^{\infty} dr \sinh r K^{{\mathbb H}_3/\Z}(t,r) = {e^{-(m^2+1)t-{(2\pi n\tau_2)^2\over 4 t}}\over
4 \pi^{3\over 2}\sqrt{t}}\,.
\end{equation}
We are left with
\begin{equation}
\begin{aligned}
&-\log \det \Delta = 2 \sum_{n=1}^{\infty} \frac{(2 \pi
\tau_2) (2\pi)}{4|\sin \pi n \tau|^2} \int_0^{\infty} {dt\over t}
{e^{-(m^2+1)t-{(2\pi n\tau_2)^2\over 4 t}}\over
4 \pi^{3\over 2}\sqrt{t}} \\
&\qquad \qquad = \sum_{n=1}^{\infty} \frac{e^{-2\pi n
\tau_2\sqrt{1+m^2}}}{2 n |\sin \pi n \tau|^2}\\
&\qquad \qquad =2 \sum_{n=1}^{\infty}
\frac{|q|^{2nh}}{n |1-q^{n}|^2}\,.
\end{aligned}
\end{equation}
where we define $h=\frac{1}{2}(1+\sqrt{1+m^2})$.

Using this, the one-loop partition function can be put in the suggestive form
\begin{equation}\label{scalarpart}
\begin{aligned}
Z_{\mbox{\tiny{scalar}}}^{\mbox{\tiny{1-loop}}}(\tau,\bar\tau)&=(\det
\Delta)^{-1/2}
=\exp\left(
\sum _{n=1}^{\infty} \frac{|q|^{2nh}}{n |1-q^{n}|^2}\right)\\
&=\exp\left(\sum_{n=1}^{\infty} \sum_{\ell,\ell'=0}^{\infty}\frac{1}{n} q^{n (\ell+h)}
\bar q^{n (\ell'+h)}\right)\\ & =\prod_{\ell,\ell'=0}^{\infty} \frac{1}{1-q^{\ell+h} \bar
q^{\ell'+h}}\,.
\end{aligned}
\end{equation}

This formula has a very natural interpretation.  We are computing a canonical ensemble partition function, so the answer should take the form of a trace
\begin{equation}\label{Ztrace-sc}
Z = \Tr q^{L_{0}}\bar q^{{\bar L}_{0}}
\end{equation}
where $L_{0}$ and $\bar L_{0}$ are related to energy and angular momentum\begin{equation}
L_{0}=H+iJ,~~~~~\bar L_{0}=H-iJ
\end{equation} In the language of the boundary conformal field theory, $L_{0}$ and $\bar L_{0}$ are Virasoro operators generating scale transformations.
Our answer (\ref{scalarpart}) is clearly of the form (\ref{Ztrace-sc}).  In fact, it is easy to understand which states are contributing to the trace (\ref{Ztrace-sc}).
Let us first examine the term in the sum (\ref{scalarpart}) with $\ell=\ell'=0$.
In the AdS/CFT correspondence, the scalar field $\phi$ is dual to a primary operator of weight $(h,h)$ in the dual conformal field theory \cite{Maldacena:1998bw}.
There is a single-particle state $|\phi\rangle$ found by inserting one of these operators at the origin.  There are also multi-particle states corresponding to multiple insertions of this operator.  The term in the sum (\ref{scalarpart}) with $\ell=\ell'=0$ is precisely the trace over these multi-particle states.  One can also construct the Virasoro descendent $L_{-1}^\ell
\bar L_{-1}^{\ell'}|\phi\rangle$ of the single particle state, where $\ell,\ell'\ge0$; this state has conformal weight $(\ell+h,\ell'+h)$.   The sum over multiple insertions of these states gives the partiton function (\ref{scalarpart}).
One might expect higher Virasoro $L_{-n}$, $n >1$ to contribute to the partition sum as well.  Indeed, we will see below that this is the case, but only once one includes the graviton one loop determinant.

\subsection{Vector fields}

The heat-kernel for a $U(1)$ gauge field on ${\mathbb H}_3/\Z$ can be
obtained from that on ${\mathbb H}_3$ by the method of images
\begin{equation}
K_{\mu\nu'}^{{\mathbb H}_3/\Z}(t,x,x') = \sum_{n \in \mathbb{Z}}
\frac{\partial (\gamma^n x)^{\rho'}}{\partial x^{\nu'}} K_{\mu\rho'}^{{\mathbb H}_3}\left(t,r(x,\gamma^n
x')\right)\,.
\end{equation}
The one loop determinant is
\begin{equation}
-\log \det {\Delta_\mu}^{\nu} = \int_0^{\infty} \frac{dt}{t}
\sum_{n \in \mathbb{Z}} \int d^3 x \sqrt{g} \,\hat g^{\mu\nu'}
 K^{H_3}_{\mu\nu'}(t,r(x,\gamma^n x))\,.
\label{gauge-logdet}
\end{equation}
where $\hat g^{\mu\nu'}\equiv g^{\mu\rho}(x) \frac{\partial (\gamma^n x)^{\nu'}}{\partial x^{\rho}}$.
We will define
\begin{eqnarray}
&&A_\gamma(r):=\hat g^{\mu\rho}(z)
\partial_\mu \partial_{\nu'} u = \cosh r-2 \cosh(2\pi \tau_2)-2 \cos(2\pi \tau_1), \nonumber\\
&&B_\gamma(r):=\hat g^{\mu\rho}(z)
\partial_\mu u \partial_{\nu'} u = (\cosh r- e^{2\pi \tau_2})
(\cosh r-e^{-2\pi \tau_2})\\
&&~~~-2 \cos(2\pi\tau_1) (\cosh r-\cosh(2\pi \tau_2)) .\nonumber
\end{eqnarray}
Using the identities in Appendix A, and making the
change of variables (\ref{change}), we get 
\begin{equation}
\begin{aligned}
&-\log \det \Delta^{\mu}_{\nu}= -3\, \mbox{vol}(H_3 / \Gamma)
\int_0^{\infty} \frac{dt}{t} \Big(F+\partial_u S
\Big)\Big{|}_{u=0} \\
& ~~~+ 2 \int_0^{\infty} \frac{dt}{t} \sum_{n=1}^{\infty} \frac{(2
\pi \tau_2) (2\pi)}{4|\sin \pi n \tau|^2} \int_{2\pi n
\tau_2}^{\infty}\!\!\!\!\!\!dr\, \sinh r \Big[A_{\gamma^n}(r) (F+
\partial_u S) + B_{\gamma^n}(r) \partial_u^2 S \Big]\,.
\label{gauge-logdet2}
\end{aligned}
\end{equation}
One may now proceed as for the scalar by first integrating over
$r$,
\begin{equation}
\int_{2\pi n
\tau_2}^{\infty}\!\!\!\!\!\!dr\, \sinh r \Big[A_{\gamma^n}(r) (F+
\partial_u S) + B_{\gamma^n}(r) \partial_u^2 S \Big] = {
e^{-{(2\pi n\tau_2)^2\over 4t}}
\over 4\pi^{3\over 2}\sqrt{t}}\left[2\cos(2\pi n\tau_1)+e^{-t}\right]
\end{equation}
The first term in the last line is the contribution of a massless
free scalar field, and the rest is the contribution of the transverse components
of the vector field. This result can be reproduced by applying the Selberg trace formula to the vector Laplacian.

The final result can be written as
\begin{equation}\label{resvec}
\begin{aligned}
-\frac{1}{2}\ln \det \Delta_\mu^\nu &=
\sum_{n=1}^\infty \frac{2\cos(2\pi n\tau_1)+e^{-2\pi n\tau_2} }{4n|\sin \pi n\tau|^2} \\
&= \sum_{n=1}^\infty \frac{q^n+\bar q^n+|q|^{2n}}{n|1-q^n|^2}
\end{aligned}
\end{equation}

The partition function of the transverse vector field from (\ref{resvec}) can be rewritten in the more suggestive form:
\begin{equation}
\begin{aligned}
Z_{\perp}^{\mbox{\tiny{1-loop}}}(\tau,\bar\tau)& = \exp\left(\sum_{n=1}^\infty \frac{q^n+\bar q^n}{n|1-q^n|^2}\right) \\
&= \prod_{\ell,\ell'=0}^\infty \frac{1}{(1-q^{\ell+1}\bar q^{\ell'})
(1-q^{\ell}\bar q^{\ell'+1})}\,.
\end{aligned}
\end{equation}
The $U(1)$ gauge field partition function is given by
\begin{equation}
Z_{\mbox{\tiny{gauge}}}^{\mbox{\tiny{1-loop}}}(\tau,\bar\tau) = {Z_{\perp}^{\mbox{\tiny{1-loop}}}(\tau,\bar\tau)\over Z_{\mbox{\tiny{scalar}}}^{\mbox{\tiny{1-loop}}}(\tau,\bar\tau)}\,.
\end{equation}
As in the scalar field case described above, these formulas have a natural boundary interpretation.

\subsection{Gravity}

The one-loop partition function of gravity is given by
\begin{equation}
Z_{\mbox{\tiny{gravity}}}^{\mbox{\tiny{1-loop}}} =
{\det \Delta^{(1)}\over
\sqrt{\det \Delta^{(2)}\det \Delta^{(0)}} }
\end{equation}
where $\Delta^{(2)}$, $\Delta^{(1)}$ and $\Delta^{(0)}$ are the kinetic operators
for the traceless symmetric tensor $\phi_{\mu\nu}$, the vector ghost $\eta_\mu$
and the Weyl mode $\phi$ as in (\ref{phigravv}), (\ref{ghostact}).

The one loop determinant for the symmetric traceless tensor $\phi_{\mu\nu}$ is
\begin{eqnarray}
&&-\log \det \Delta^{(2)} = \int_0^{\infty} \frac{dt}{t}
\sum_{n \in \mathbb{Z}} \int d^3 x \sqrt{g} \,
\hat g^{\mu\mu'} \hat g^{\nu \nu'}K^{{\mathbb H}_3}_{\mu\nu,\mu'\nu'}(t,r(x,\gamma^n x))\nonumber\\
&&= \int_0^\infty {dt\over t}
\sum_{n=1}^\infty {2\pi^2\tau_2\over |\sin (\pi n\tau)|^2} \int_{2\pi n\tau_2}^\infty dr \sinh r\,
\hat g^{\mu\mu'} \hat g^{\nu \nu'}K^{{\mathbb H}_3}_{\mu\nu,\mu'\nu'}(t,r(x,\gamma^n x))\label{grav-logdet}
\end{eqnarray}
where we have omitted the term proportional to the volume of ${\mathbb H}_3/\Z$ as before.
Define
\begin{eqnarray}
&&C_\gamma(r) = \hat g^{\mu\nu'} \hat g^{\nu\nu'} g_{\mu'\nu'} \partial_\mu u \partial_\nu u = \sinh^2 r, \nonumber\\
&&J_\gamma(r) = \hat g^{\mu\mu'} \hat g^{\nu\nu'} \partial_\mu u \partial_{\nu'} u \partial_{\mu'}\partial_{\nu} u \nonumber\\
&&~~~ = (\cosh r-e^{2\pi \tau_2}) (\cosh r - e^{-2\pi \tau_2})(\cosh r - 2 \cosh (2\pi\tau_2))\nonumber\\
&&~~~~~+ 2(\cosh r - \cosh (2\pi\tau_2)) \left[\cos(4\pi\tau_1)-2 \cos(2\pi\tau_1) (\cosh r - \cosh(2\pi\tau_2)) \right],\\
&& L_\gamma(r) = \hat g^{\mu\mu'} \hat g^{\nu\nu'} \partial_\mu \partial_{\nu'} u \partial_\nu \partial_{\mu'} u \nonumber\\
&& ~~~= (\cosh r - 2 \cosh (2\pi\tau_2))^2 - 4 \cos(2\pi \tau_1) (\cosh r - \cosh (2\pi\tau_2)) + 2 \cos(4\pi\tau_1).\nonumber
\end{eqnarray}
We can then express
\begin{eqnarray}
&&\hat g^{\mu\mu'} \hat g^{\nu \nu'}K^{H_3}_{\mu\nu,\mu'\nu'}(t,r(z,\gamma z)) =
(A_\gamma^2 + L_\gamma) (G+X) + 3 H + 6(u+1) Z \nonumber\\
&&~~~+ (A_\gamma B_\gamma + J_\gamma) (\partial_u X + 2Y)
+ 2 C_\gamma (X+(u+1) Y+\partial_u Z) + 2 B_\gamma^2 \partial_u Y,
\end{eqnarray}
where $u+1=\cosh r$.
With some effort, the integral over $r$ in (\ref{grav-logdet}) can be evaluated, and the result is
\begin{eqnarray}
&&\int_{2\pi n\tau_2}^\infty dr \sinh r\,
\hat g^{\mu\mu'} \hat g^{\nu \nu'}K^{{\mathbb H}_3}_{\mu\nu,\mu'\nu'}(t,r(z,\gamma^n z)) \nonumber\\
&&={e^{-{(2\pi n\tau_2)^2\over 4t}}\over 2\pi^{3\over 2} \sqrt{t}} \left[
e^{-t} \cos(4\pi n\tau_1) + e^{-4t}\cos(2\pi n\tau_1) +{e^{-5 t}\over 2} \right].
\end{eqnarray}
The contribution from the vector ghost and the Weyl mode can be obtained analogously to the vector
and scalar field cases discussed before (with appropriate mass terms)
\begin{eqnarray}
-{e^{-{(2\pi \tau_2)^2\over 4t}}\over 2\pi^{3\over 2} \sqrt{t}} \left[
2 e^{-4t}\cos(2\pi\tau_1) +e^{-5 t} \right]+{e^{-{(2\pi \tau_2)^2\over 4t}-5t}\over 4\pi^{3\over 2} \sqrt{t}}.
\end{eqnarray}
The full 1-loop free energy is given by
\begin{eqnarray}
\ln Z_{\mbox{\tiny{gravity}}}^{\mbox{\tiny{1-loop}}} &=& -{1\over 2} \ln \det \Delta^{(2)} + \ln \det \Delta^{(1)} - {1\over 2}\ln\det \Delta^{(0)}
\nonumber\\
&=& \int_0^\infty {dt\over t} \sum_{n=1}^\infty {2\pi^2\tau_2 \over |\sin (\pi n\tau)|^2}
{e^{-{(2\pi n\tau_2)^2\over 4t}}\over 4\pi^{3\over 2} \sqrt{t}} \left[
e^{-t} \cos(4\pi n\tau_1) -e^{-4t}\cos(2\pi n\tau_1) \right] \nonumber\\
&=& \sum_{n=1}^\infty {q^{2n}+\bar q^{2n} - |q|^{2n} (q^n+\bar q^n)\over n |1-q^n|^2} \nonumber\\
&=& \sum_{n=1}^\infty {1\over n}\left({q^{2n}\over 1-q^n}+{\bar q^{2n}\over 1-\bar q^n}\right)\nonumber\\
&=& -\sum_{m=2}^\infty \ln |1-q^{m}|^2.
\end{eqnarray}
Or,
\begin{equation}
Z_{\mbox{\tiny{gravity}}}^{\mbox{\tiny{1-loop}}}(\tau,\bar\tau) = \prod_{m=2}^\infty {1\over |1-q^m|^2}.\label{solidgrav}
\end{equation}

This one loop contribution is in addition to the tree level gravity partition function, which is found by computing the regularized volume of $\bH_{3}/\Z$. This tree level contribution is $|q|^{-2k} $, as described in e.g. \cite{Maldacena:1998bw}.  The full gravity partition function is therefore
\begin{equation}Z_{\mbox{\tiny{gravity}}}(\tau,\bar \tau)= |q|^{-2k} \prod_{m=2}^{\infty}{1\over |1-q^{m}|^{2}}\label{finalans}\end{equation}
This formula has a very natural physical interpretation.  It has the form of a trace     $$Z=\Tr q^{L_{0}} {\bar q}^{{\bar L}_{0}}$$ over an irreducible representation of the Virasoro algebra.  This representation contains a ground state $|0\rangle$ of weight $L_{0}|0\rangle = -k|0\rangle$, along with its Virasoro descendants $L_{-n_{1}}\dots L_{-n_{i}} |0\rangle$.  This result is not surprising given the observation of Brown \& Henneaux \cite{Brown:1986nw} that, with appropriate boundary conditions, the symmetry group relevant to AdS$_{3}$ gravity is generated by the Virasoro algebra.  Our computation may therefore be viewed as  an explicit check that quantum gravity in AdS$_{3}$ does indeed have the structure of a conformal field theory.  A detailed derivation of the partition function (\ref{finalans}) from the Brown-Henneaux construction was given in \cite{Maloney:2007ud}.  As described in \cite{Maloney:2007ud}, these symmetry arguments indicate that the expression (\ref{finalans}) for the gravity partition function is in fact one-loop exact.

\section{Hyperbolic manifolds with higher genus conformal boundary}

In this section we will consider more general quotients of $\bH_{3}$ by some discrete subgroup $\Gamma$ of the isometry group $PSL(2,\bC)$ of $\bH_{3}$.  We will use the heat kernel of section 2 to compute the one-loop contribution to the gravity partition function, generalizing the results for $\bH_{3}/\bZ$ described in the previous section. The goal of these computations is to compute the one-loop partition function of gravity on a hyperbolic 3-manifold whose conformal boundary is a genus $g$ Riemann surface. We will start by describing a few salient features of the quotients $H_{3}/\Gamma$.  

\subsection{The Geometry of $\bH_{3}/\Gamma$}

We are interested in quotients $\bH_{3}/\Gamma$ whose conformal boundary is a Riemann surface of genus $g$.  Each such geometry gives a saddle point contribution to the partition function of the quantum gravity path integral at fixed genus. 

We will start by considering 3-manifolds whose boundary is a genus $1$ surface, $T^{2}$.  There are two types of smooth hyperbolic manifolds with $T^{2}$ boundary.  The first is the solid torus $\bH_{3}/\bZ$ described in section 4.  The second is the quotient $\bH_{3}/\Z\times\Z$.  In terms of the $(y,z)$ coordinates of equation (\ref{H3-metric}), $\bH_{3}/\Z\times\Z$ is given by the identifications $z\sim z+1 \sim z+\tau$.  From the form of the metric (\ref{H3-metric}) it is apparent that $\bH_{3}/\Z\times\Z$ has a $T^{2}$ conformal boundary at $y=0$, as well as a cusp at $y\to \infty$.  In fact, this cusp renders this geometry unstable; one can show, by computing the heat-kernel on $\bH_{3}/\Z\times\Z$, that the spectrum of linearized metric perturbations on $\bH_{3}/\Z\times\Z$ has a negative mode.  For this reason, we will not consider geometries with cusps in what follows.

The geometries $\bH_{3}/\Gamma$ with conformal boundary of genus $g\ge 2$ are considerably more complicated.  The simplest class of
such manifolds are genus $g$ handlebodies endowed with metrics of constant negative curvature.  In this case the group $\Gamma$ is freely generated by $g$ loxodromic elements of $PSL(2,\bC)$.  These handlebodies may be thought of as the higher genus generatizations of the BTZ black hole \cite{Aminneborg:1997pz, Krasnov:2000zq}. These 3-manifolds often have interesting Lorentzian continuations, which describe multiple asymptotic AdS regions connected by wormholes of complicated topology. 

There are also non-handlebodies 3-manifolds, which have genus $g\ge2$ conformal boundary.  The geometry of these 3-manifolds is somewhat more complex (see \cite{maskit, Yin:2007at} for some examples).  These non-handlebodies also have interesting Lorentzian continuations, which include the closed FRW cosmologies of \cite{Horowitz:1998xk, Maldacena:2004rf}.

In this section we will consider smooth hyperbolic manifolds $M={\mathbb H}_3/\Gamma$
without cusps.  This means that the group $\Gamma$ will contain only loxodromic elements, and will not have
a subgroup isomorphic to ${\mathbb Z}\times {\mathbb Z}$.

\subsection{Heat Kernels}

We will now compute the heat kernel of a free field on ${\mathbb H}_3/\Gamma$.  Using the method of images, this is related to the heat kernel of a free field on $\bH_{3}$ by
\begin{equation}
K^{{\mathbb H}_3/\Gamma}(t,x,x') = \sum_{\gamma\in\Gamma} K^{{\mathbb H}_3}(t,r(x,\gamma x'))
\end{equation}
This formula applies to scalar, gauge or graviton fields (for the latter two cases we have omitted the index structure, see sections 4.3 and 4.4 for details).  We will denote by ${\cal F}$ a fundamental region for $\Gamma$ on $\bH_{3}$.  Then the one-loop determinant is given by
\begin{eqnarray}
-\ln \det \Delta &=& \int_0^\infty {dt\over t} \sum_{\gamma\in\Gamma} \int_{\cal F} d^3x \sqrt{g} K^{{\mathbb H}_3}(t,r(x,\gamma x))
\nonumber\\
&=& \int_0^\infty {dt\over t} \sum_{\gamma\in {\cal P}} \sum_{\sigma \in {\cal I}_\gamma} \sum_{n=1}^\infty
\int_{\cal F} d^3x \sqrt{g} K^{{\mathbb H}_3}(t,r(x,\sigma^{-1}\gamma^n \sigma x))\nonumber
\end{eqnarray}
In the second line we have separated the sum over $\Gamma$ into three parts.   Here ${\cal I}_\gamma$ is the set of $\sigma\in \Gamma$ such that
$\sigma^{-1}\gamma\sigma$ give all the distinct elements in the conjugacy class of $\gamma$, and ${\cal P}$ denotes a set of representatives of the primitive conjugacy classes of $\Gamma$ (an element $\gamma\in \Gamma$ is primitive if $\gamma\not=\beta^n$ for any
any element $\beta\in\Gamma$ and $n>1$). 
We may rewrite this as 
\begin{eqnarray}
-\ln \det \Delta
&=& \int_0^\infty {dt\over t} \sum_{\gamma\in {\cal P}} \sum_{\sigma \in {\cal I}_\gamma}
\sum_{n=1}^\infty \int_{\sigma \cal F} d^3x \sqrt{g} K^{{\mathbb H}_3}(t,r(x,\gamma^n x))\nonumber \\
&=& \int_0^\infty {dt\over t} \sum_{\gamma\in {\cal P}} \sum_{n=1}^\infty
\int_{\cal F_\gamma} d^3x \sqrt{g} K^{{\mathbb H}_3}(t,r(x,\gamma^n x)) \label{conj}
\end{eqnarray}
where ${\cal F}_\gamma$ is a fundamental domain for the group $\bZ$ generated by the element $\gamma$. In the last step
of (\ref{conj}), we have used the fact that $\bigcup_{\sigma\in{\cal I}_\gamma} \sigma{\cal F}$ is a fundamental domain
for $\gamma$.\footnote{To see this, we need to show that (1) for any $x\in \bigcup_{\sigma\in{\cal I}_\gamma}\sigma{\cal F}$ and $n\not=0$,
$\gamma^n x\not\in \bigcup_{\sigma\in{\cal I}_\gamma}\sigma{\cal F}$, and (2) for any point $y\in {\mathbb H}_3$, there exists some
$x\in \bigcup_{\sigma\in{\cal I}_\gamma}\sigma{\cal F}$ and $n\in {\mathbb Z}$ such that $y=\gamma^n x$.
(1) follows from the definition of ${\cal I}_\gamma$. To show (2), first note that there is some $\alpha\in {\cal F}$ such that
$y=g\alpha$ for an element $g\in\Gamma$. It follows from the definition of ${\cal I}_\gamma$ that
$g^{-1}\gamma g=\sigma^{-1}\gamma\sigma$ for some $\sigma\in {\cal I}_\gamma$.
Equivalently, $g\sigma^{-1}$ commutes with $\gamma$. Since $\Gamma$ does not contain $\Z\times \Z$, it follows that $g\sigma^{-1}=\gamma^n$
for some $n$, i.e. $y=\gamma^n\sigma\alpha=\gamma^n x$.
} As in section 4, we have omitted the term proportional to the volume of ${\mathbb H}_3/\Gamma$, corresponding
to $\gamma={\bf 1}$.  This term describes the renormalization of the cosmological constant at one loop.

We may now use the fact that, aside from the sum over $\cal P$, equation (\ref{conj}) is identical to equation (\ref{scalar-logdet}) for the partition function on $\bH_{3}/\Z$.  The implies that the one-loop partition function on $\bH_{3}/\Gamma$ is given by
\begin{eqnarray}
Z^{{\mathbb H}_3/\Gamma} &=& \prod_{\gamma\in {\cal P}} \left(Z^{{\mathbb H}_3/\langle \gamma \rangle}\right)^{1\over 2}
\end{eqnarray}
Here $Z^{{\mathbb H}_3/\langle \gamma \rangle}$ is the one-loop partition function on the solid torus $\bH_{3}/\Z$, where $\bZ=\langle\gamma\rangle$ is the free group generated by $\gamma$.  To evaluate $Z^{\bH/\langle\gamma\rangle }$ explicitly, it is useful to use a basis where the element $\gamma\in PSL(2,\bC)$ is diagonal: $\gamma = \left(\begin{array}{cc}
q_\gamma^{1/ 2} & 0 \\ 0 & q_\gamma^{-{1/ 2}}\end{array}\right)$, with $|q_{\gamma}| < 1$.  Then $Z^{\bH/\langle\gamma\rangle}$ is just the one-loop partition function $Z^{\bH_{3}/\Z}$
computed in section 4, evaluated at the point $q_\gamma$. 

This argument applies
to vector and graviton fields as well as to scalar fields. In the scalar and vector cases, our result can be reproduced by the Selberg
trace formula applied to the scalar and vector Laplacians. In the gravity case, we arrive at the following formula for
the one-loop partition function on ${\mathbb H}_3/\Gamma$,
\begin{equation}
Z^{{\mathbb H}_3/\Gamma}_{\mbox{\tiny gravity}} = \prod_{\gamma\in{\cal P}} \prod_{m=2}^\infty
{1\over |1-q_\gamma^m|}. \label{gengrav}
\end{equation}
This is precisely the expression anticipated in \cite{Yin:2007gv}. When $M={\mathbb H}_3/\Gamma$ is a solid torus, $\Gamma\simeq {\mathbb Z}$, and ${\cal P}$ consists of the generator of
$\Gamma$ and its inverse.  In this case (\ref{gengrav}) reproduces (\ref{solidgrav}).

As an example of the application of this formula, it is instructive to consider the case where $\Gamma$ is a Fuchsian group, i.e.
$\Gamma\subset SL(2,{\mathbb R})\subset SL(2,{\mathbb C})$. In this case $M={\mathbb H}_3/\Gamma$ has two conformal boundary components,
$\Sigma$ and $\overline{\Sigma}$, which are Riemann surfaces with opposite complex structures.\footnote{When $\Sigma$
admits an anti-holomorphic fixed point free involution, we can further take the quotient of $M$ by a ${\mathbb Z}_2$ to
obtain a non-handlebody with a single connected conformal boundary.} In this case, the gravity partition function can be written as
\begin{equation}
Z^{{\mathbb H}_3/\Gamma}_{\mbox{\tiny gravity}} = \prod_{\gamma\in{\cal P}} \prod_{m=2}^\infty
{1\over 1-e^{-m \ell(\gamma)}}.
\end{equation}
where ${\cal P}$ can be equivalently thought of as the set of oriented primitive geodesics on
the Riemann surface $\Sigma$ (with hyperbolic metric), and $\ell(\gamma)$ is the length of the geodesic corresponding
to $\gamma$. It is well known that the number of primitive geodesics of length $\leq L$ grows like $e^L/L$ (see e.g. \cite{Bogomolny}). It follows that
\begin{equation}
\begin{aligned}
\ln Z^{{\mathbb H}_3/\Gamma}_{\mbox{\tiny gravity}} &= \sum_{\gamma\in{\cal P}} \sum_{m=2}^\infty \ln (1- e^{-m \ell(\gamma)})
\\
&< C_1 \int_{L_0}^\infty dL {e^L\over L} \sum_{m=2}^\infty \ln (1- e^{-m L}) + C_2 <\infty
\end{aligned}
\end{equation}
for some positive constants $C_1$, $C_2$, $L_0$, i.e. the infinite product in (\ref{gengrav}) converges.\footnote{Note that
had the product over $m$ in (\ref{gengrav}) started from $m=1$ (analogously to the formula for the classical regularized action of a handlebody
in \cite{Yin:2007gv, Yin:2007at}), the product over $\gamma\in {\cal P}$ would diverge for Fuchsian $\Gamma$.}

\

It is worth noting that the higher genus partition functions discussed above encode in principle the correlation functions of all operators in the boundary CFT.  Unlike the $\bH_{3}/\bZ$ case described in section 4, there is no reason why higher-loop contributions to these partition function should vanish. So we expect that the results described above will be corrected at higher loop order.  When $\cM$ is a handlebody, a method to obtain the exact all-loop expression has been proposed in \cite{Yin:2007gv} (and explicit results were given for genus two). It would be interesting if these higher loop contributions could be computed explicitly in gravity perturbation theory. 

\subsection*{Acknowledgments}

The authors would like to thank D. Freedman, J. Maldacena and especially E. Witten for useful conversations.
The work of S.G. is supported in part by NSF grants PHY-024482 and DMS-0244464. The work of A.M. is supported by the National Science and Engineering Research Council of Canada.
The work of X.Y. is supported by a Junior Fellowship from the Harvard Society of Fellows.

\appendix

\section{Bitensor Identities on $\bH_{3}$}

In this appendix we collect a few useful identities that are used in the derivation of heat kernels on $\bH_{3}$.  Reference \cite{Freedman} contains a more lengthy discussion of many of these identities, which were used to construct  gauge and graviton two point functions in Anti-de Sitter space.

We start by noting that the chordal distance $u(x,x')$ defined in equation (\ref{chordal})
obeys
\bea
\p^{\m}\p_{\m}u=&3(u+1)\cr
\p^{\m}u\p_{\m}u=&u(u+2)\cr
\p_{\m}\p_{\n}u=&g_{\m\n}(1+u)\cr
\p^{\m}u \p_{\m}\p_{\n}\p_{\n'}u=&\p_{\n}u\p_{\n'}u\cr
\p^{\m}u\p_{\m}\p_{\n'}u=&(1+u)\p_{\n'}u\cr
\p^{\m}\p_{\m'}u \p_{\m}\p_{\n'}u=&g_{\m'\n'}+\p_{\m'}u\p_{\n'}u\cr
\p_{\m}\p_{\n}\p_{\n'}u=&g_{\m\n}\p_{\n'}u
\eea

The gauge and graviton heat kernel are bitensors constructed out of $u(x,x')$.  There are two linearly independent $(1,1)$ bitensors that can be constructed out of $u$:
\begin{equation}
W^{1}_{\mu\mu'}=\p_{\mu} u \p_{\mu'} u\,,~~~~~~
W^{2}_{\mu\mu'}=\p_{\mu}\p_{\mu'} u
\end{equation}

The gauge field heat kernel can be written as a linear combination
\begin{equation}
K_{\mu\nu'}(t,x,x') = \sum_{a=1}^{2}K_{a}(t,u) W^{a}_{\mu\nu'}
\end{equation}
However, the computations are considerably simpler if one writes the heat kernel in the form (\ref{heataa}) instead.  It is straightforward to demonstrate that the decomposition (\ref{heataa}) is equivalent to the form described above; the $F$ and $S$ of equation (\ref{heataa}) are uniquely determined by the $W^{a}$ defined above.
In writing down the heat kernel equation (\ref{gaugeheat}) we use the identities
\begin{equation}
\begin{aligned}
& [\nabla^2, \nabla_\mu] f = -2 \nabla_\mu f \,, \\
& [\nabla^2, \nabla_\mu] \upsilon_\nu = 2g_{\mu\nu}\nabla^\rho \upsilon_\rho -
2 \nabla_\mu \upsilon_\nu -2 \nabla_\nu \upsilon_\mu \,,
\end{aligned}
\label{commutators}
\end{equation}
for a scalar function $f$ and a vector $\upsilon_\mu$.

To study the graviton propagator $K_{\mu\nu,\mu'\nu'}$, we note that one can construct six, linearly independent rank-four, symmetric bilinear tensors from $u(x,x')$
\bea
T^{1}_{\m\n\m'\n'}&=&g_{\m\n}g_{\m'\n'}
\cr
T^{2}_{\m\n\m'\n'}&=&\p_{\m}u\p_{\n}u\p_{\m'}u\p_{\n'}u
\cr
T^{3}_{\m\n\m'\n'}&=&\p_{\m}\p_{(\m'}u\p_{\n')}\p_{\n}u
\cr
T^{4}_{\m\n\m'\n'}&=&\p_{(\n}u \p_{\m)}\p_{(\m'}u \p_{\n')}u
\cr
T^{5}_{\m\n\m'\n'}&=&g_{\m\n}\p_{\m'}u\p_{\n'}u
\cr
T^{6}_{\m\n\m'\n'}&=&\p_{\m}u\p_{\n}u g_{\m'\n'}
\eea
These are all symmetric under $\m\leftrightarrow \n$ and $\m'\leftrightarrow\n'$.  Under $x\leftrightarrow x'$, the first four tensors are invariant, while
$T^{5} \leftrightarrow T^{6}$.  The first four tensors, along with $T^{5}+T^{6}$, form a linearly independent basis of the space of symmetric, bilinear tensors invariant under $x\leftrightarrow x'$.
We may therefore write the graviton heat kernel as the sum
\be
K_{\m\n\m'\n'}^{{\mathbb H}_{3}} = \sum_{i=1}^{6}K_{i}T^{i}_{\m\n\m'\n'}
\ee
where $i$ runs from $1$ to $6$.  Since the heat kernel is invariant under $x\to x'$ it follows that $K_{5}=K_{6}$.
However, the computations are considerably simpler if one writes the heat kernel in the form (\ref{kexpansion}) instead.  It is straightforward but tedious to demonstrate that the decomposition (\ref{kexpansion}) is equivalent to the form described above.
In writing down the equation of motion (\ref{gravitoneom}) we again used the identities (\ref{commutators}).

\end{document}